\documentclass[aps,prb,twocolumn,groupedaddress,floatfix,showpacs,amsmath,amssymb]{revtex4} 
\usepackage{graphicx}
\usepackage{amsmath}
\usepackage{psfrag}
\begin{document}
\newcommand\ket[1]{|#1\rangle}
\newcommand\bra[1]{\langle#1|}
\newcommand\braket[2]{\left\langle#1\left|#2\right.\right\rangle}

\title{Adiabatic charge and spin pumping through quantum dots with
  ferromagnetic leads}

\author{Janine Splettstoesser}
\affiliation{D\'epartement de Physique Th\'eorique, Universit\'e de Gen\`eve,
 CH-1211 Gen\`eve 4, Switzerland}
\author{Michele Governale}
\affiliation{Institut f\"ur Theoretische Physik III,
Ruhr-Universit\"at Bochum, D-44780 Bochum, Germany}
\author{J\"urgen K\"onig}
\affiliation{Institut f\"ur Theoretische Physik III,
Ruhr-Universit\"at Bochum, D-44780 Bochum, Germany}

\date{\today}

\begin{abstract}
We study adiabatic pumping of electrons through quantum dots attached to 
ferromagnetic leads. Hereby we make use of a real-time diagrammatic technique
in  the adiabatic limit that takes into account strong Coulomb interaction in
the dot.
We analyze the degree of spin polarization of electrons pumped from a
ferromagnet through the dot to a nonmagnetic lead (N-dot-F) as well as the
dependence of the pumped charge on the relative leads' magnetization
orientations for a spin-valve (F-dot-F) structure.
For the former case, we find that, depending on the relative coupling
strength to the leads, spin and charge can, on average, be pumped in opposite
directions. 
For the latter case, we find an angular dependence of the pumped charge,
that becomes more and more anharmonic for large spin polarization in the leads.
\end{abstract}


\maketitle

\section{Introduction}

Charge and spin transport through a nanoscale conductor can be obtained, 
in the absence of a transport voltage,  by periodically varying in time some
of its parameters.  
If the time dependence of the system is slow compared to its characteristic
response time, we refer to this transport mechanism as adiabatic pumping. 
This particular regime allows us to study the properties of a system being  
slightly out of  equilibrium due to an explicit time-dependence of its
parameters. Numerous works have studied mesoscopic pumps both theoretically 
\cite{brouwer98,zhou99,buttiker01,buttiker02,entin02} as well as
experimentally.\cite{switkes99,pothier92,geerligs91,fletcher03,watson03} 
The established framework to calculate the pumped charge through a mesoscopic
scatterer is based on the dynamical scattering
approach.\cite{buttiker94,brouwer98} This approach can be applied when the
Coulomb interaction can be neglected or treated within the Hartree
approximation.   
Recently, the interest in including the effects of Coulomb interaction beyond
the Hartree  level to the problem of adiabatic pumping has
arisen.\cite{aleiner98,citro03,aono04,brouwer05,cota05,splett05,sela06,
splett06,fioretto07}   

\begin{figure}
\begin{center}
\includegraphics[width=3.3in]{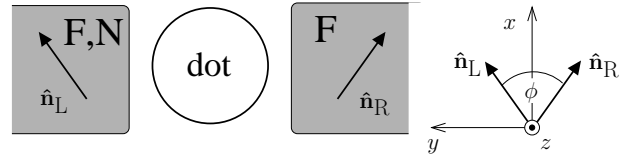}
\caption{
  Schematic setup of N-dot-F or F-dot-F setup. The magnetization
  directions and the polarization strengths of the left and right lead can in
  general differ from each other (left). Sketch of the coordinate system and
  the magnetization directions of the leads (right).}
\label{fig_fdotf}
\end{center}
\end{figure}

Spin-dependent transport through nanostructures has recently attracted a lot of
interest.
A model example is a quantum-dot spin valve, which consists
of an interacting (single-level) quantum dot attached to two 
ferromagnetic leads (F-dot-F), see Fig. \ref{fig_fdotf}. The leads have, in
general, non-collinear magnetization directions and different polarization
strengths.  
Transport through a quantum-dot spin valve with noncollinear leads has been
studied extensively in the dc limit.\cite{braun04,QDSV}  
In magnetic multilayers the tunneling current depends chiefly on the relative
orientation of the magnetization of the ferromagnetic
layers.\cite{julliere75,slonczewski} The situation is more complex in a
quantum-dot spin valve due to the interplay of the lead magnetization, Coulomb
interaction, non-equilibrium spin accumulation, and quantum fluctuations. 
In particular, a finite spin accumulation is generated on the dot, which plays 
an important role in determining charge transport.

In the present paper we combine the ideas of adiabatic pumping and
spin-dependent transport through interacting nanostructures.
We consider two scenarios.
First, we focus on the situation when only one of the two leads is
ferromagnetic (N-dot-F), for which we study the spin pumped into the
non-magnetic lead. 
Spin pumping in systems where the spin degeneracy is lifted by a magnetic 
field has been the subject
of several studies.\cite{mucciolo02,blaauboer03,watson03,aono04,sela06}  
Furthermore spin pumping by means of electrical gating only was predicted in a
system with Rashba spin-orbit coupling.\cite{governale03}
Several aspects of a noninteracting spin pump based on ferromagnets were
studied in Ref. \onlinecite{zheng03}. 
Spin pumping through an interface between a ferromagnet and a non-magnetic
metal has been investigated in Ref.~\onlinecite{brataas02}, where the pumping
cycle is realized exploiting the precession of the magnetization of the
ferromagnet. 
In our setup we are interested in spin pumping obtained by varying 
periodically the properties of the scattering
region, such as the dot level position and the tunneling strength to the left
and the right lead, but leaving the lead properties, as e.g. their
magnetizations, constant in time. A particular intriguing result for the
system under consideration is that, depending on the relative coupling
strength between dot and leads, spin and charge can on average be transported
in opposite directions.

Second, we consider the case that both leads are spin polarized (F-dot-F).
We study the influence of the spin polarizations of the leads on the pumped
charge and on the average spin accumulated on the quantum dot during a pumping
cycle. 
As a result, we find that also the pumped charge displays the spin-valve 
effect, i.e., a dependence on the relative angle between magnetization
direction of the leads. 
For stronger spin polarization of the leads, the pumped charge becomes a more
and more anharmonic function of the relative angle.

In order to calculate the charge and the spin pumped through the dot we
use real-time diagrammatic technique in the adiabatic limit
\cite{koenig96,splett06} and perform a rigorous perturbation expansion in the
tunnel coupling to the leads.
We consider the system in the regime of weak coupling, considering only
first-order processes in the tunnel-coupling strengths.

\section{Model and Formalism}

\subsection{Hamiltonian}

We consider a single-level quantum dot contacted by tunnel barriers to two
 ferromagnetic leads with different spin polarization axes, as shown in
 Fig. \ref{fig_fdotf}. 
For finite spin polarization in both leads (F-dot-F), the system is called a
quantum-dot spin valve. 
The limit of one normal and one ferromagnetic lead (N-dot-F) is included 
by setting the spin polarization of one lead to zero.
The total Hamiltonian of the system can be written as
\begin{equation}
H=H_{\mathrm{dot}}+\sum_{\alpha=\mathrm{L,R}}H_{\mathrm{lead, \alpha}}
+H_{\mathrm{tunnel}}\ .
\end{equation}
It consists of the Hamilton operators for the dot, for the left (L) and right
 (R) lead and for electron tunneling between dot and leads.
 The single-level quantum dot is described by the Hamiltonian
\begin{equation}
H_\mathrm{dot}=\sum_{\sigma=\uparrow,\downarrow}\epsilon_\sigma(t) n_\sigma + U
n_\uparrow n_\downarrow\ ,
\end{equation}
where the operator $d^\dagger_\sigma (d^{}_\sigma)$ creates (annihilates) an
electron with spin $\sigma=\uparrow, \downarrow$ on the dot and   $
n_\sigma=d^\dagger_\sigma d^{}_\sigma$ is the number operator for electrons
with spin $\sigma$. The strength of the Coulomb interaction between electrons
on the dot is denoted by $U$, which can be arbitrarily large.
The energy level 
$\epsilon_\sigma(t)=\bar{\epsilon}_\sigma+\delta\epsilon_\sigma(t)$ of the dot
can vary in time. In the following, we assume the dot level to be
spin degenerate, i.e.  
$\epsilon_\uparrow(t)=\epsilon_\downarrow(t)=\epsilon(t)$. 
The Hamiltonian of the lead $\alpha$, with $\alpha=\mathrm{L, R}$ is given by
\begin{equation}
H_{\mathrm{lead},\alpha}=\sum_{k\sigma}
\epsilon_{k\sigma}c^{\dagger}_{\alpha k \sigma}
c^{}_{\alpha k \sigma}\,. 
\end{equation}
We choose the spin quantization axis of lead $\alpha$ along the direction of
its magnetization,  $\mathbf{\hat{n}}_\alpha$. The spin $\sigma$ of an
electron in lead $\alpha$ can take the values $\sigma=\pm$, where $+$
refers to the majority spin and  $-$ to the minority spin of this lead. 
We choose a coordinate system with the three basis vectors
$\mathbf{\hat{e}}_x$, $\mathbf{\hat{e}}_y$ and $\mathbf{\hat{e}}_z$, 
pointing along
$\mathbf{\hat{n}}_\mathrm{L}+\mathbf{\hat{n}}_\mathrm{R}$,
$\mathbf{\hat{n}}_\mathrm{L}-\mathbf{\hat{n}}_\mathrm{R}$
and $\mathbf{\hat{n}}_\mathrm{R}\times\mathbf{\hat{n}}_\mathrm{L}$
respectively, in analogy to the definition in Ref. \onlinecite{braun04}. The
angle between the spin quantization axis of the left lead,
$\mathbf{\hat{n}}_\mathrm{L}$, and the spin quantization axis of the right
lead, $\mathbf{\hat{n}}_\mathrm{R}$, is given by  $\phi$. 
We show a sketch of the coordinate system and of the magnetization directions
of the leads in Fig. \ref{fig_fdotf}. As spin quantization axis of the dot we
take the $z$-axis of this coordinate system. With this choice the tunneling
Hamiltonian reads  
\begin{eqnarray}
H_{\mathrm{tunnel}} & = & \frac{V_\mathrm{L}(t)}{\sqrt{2}}\sum_{k}
\left[c^{\dagger}_{\mathrm{L}k+}\left(e^{i\phi/4}d^{}_{\uparrow}+
e^{-i\phi/4}d^{}_{\downarrow}\right)\right.\\
 & &
 \left.+c^{\dagger}_{\mathrm{L}k-}\left(-e^{i\phi/4}d^{}_{\uparrow}+
e^{-i\phi/4}d^{}_{\downarrow}\right)\right]\nonumber\\
& + & \frac{V_\mathrm{R}(t)}{\sqrt{2}}\sum_{k}
\left[c^{\dagger}_{\mathrm{R}k+}\left(e^{-i\phi/4}d^{}_{\uparrow}+
e^{i\phi/4}d^{}_{\downarrow}\right)\right.\nonumber\\
 & &
 \left.+c^{\dagger}_{\mathrm{R}k-}\left(-e^{-i\phi/4}d^{}_{\uparrow}
+e^{i\phi/4}d^{}_{\downarrow}\right)\right]+
 \mathrm{h.c.}\nonumber
\end{eqnarray}
The tunnel matrix elements, $V_\mathrm{L}(t)$ and
$V_\mathrm{R}(t)$, can be both time dependent. The generalized tunnel 
rates are defined by $\Gamma_\alpha\left(t,t'\right)=\frac{1}{2}
\sum_{\sigma=\pm}2\pi V^{*}_\alpha\left(t\right)
V^{}_\alpha\left(t'\right)\rho_{\alpha,\sigma}=
\frac{1}{2}\sum_{\sigma=\pm}\Gamma_{\alpha,\sigma}
\left(t,t'\right)$
and $\Gamma_\alpha\left(t\right)=\Gamma_\alpha\left(t,t\right)$. Here
$\rho_{\alpha,\sigma}$ is 
the density of states of the spin species $\sigma$ in lead
$\alpha$, which is supposed to be constant.
The spin polarization of lead $\alpha$ is defined as
\begin{equation}
p_\alpha=\frac{\rho_{\alpha+}-\rho_{\alpha-}}{\rho_{\alpha+}+\rho_{\alpha-}}\ ,
\end{equation}
and it can take values between $0$ and $1$.

\subsection{Real-time diagrammatic approach}
The Hilbert space of the single-level quantum dot is four dimensional, and it
is spanned  by the states \hbox{$\chi=0,\uparrow,\downarrow, d$} (empty dot,
singly occupied dot with spin up, singly occupied dot with spin down, doubly
occupied dot). 
On the other hand, the non-interacting leads attached to the dot have a large
number of degrees of freedom and act as baths. Hence, we can trace them out to
obtain an effective description of the quantum dot. The dot dynamics are fully
described by its reduced density 
matrix, $\rho_{\mathrm{dot}}$, with matrix elements
$P_{\chi_2}^{\chi_1}=\langle \chi_2|\rho_{\mathrm{dot}}|\chi_1\rangle$. We
introduce also the notation $P_\chi=P_{\chi}^{\chi}$  
for the diagonal matrix elements (probabilities).
The time evolution of the reduced density matrix is governed by the
generalized master equation 
\begin{eqnarray}
\label{master}
\frac{d}{dt}P^{\chi_1}_{\chi_2}\left(t\right) & = &
-\frac{i}{\hbar}(E_{\chi_1}-E_{\chi_2})P^{\chi_1}_{\chi_2}\left(t\right)\\
&&+\sum_{\chi_1',\chi_2'}\int_{-\infty}^{t}dt'W^{\chi_1\chi_1'}_{\chi_2\chi_2'}
\left(t,t'\right)P^{\chi_1'}_{\chi_2'}\left(t'\right)\ .\nonumber
\end{eqnarray}
The kernel $W^{\chi_1\chi_1'}_{\chi_2\chi_2'}\left(t,t'\right)$  connects the  
states $\chi_1'$ and  $\chi_2'$ at time $t'$ with the states $\chi_1$ and
$\chi_2$ at time $t$. It is useful to define the vector of the average
occupation probabilities $\mathbf{P}=\left(P_0,P_1,P_d\right)=
\left(P_0,P_\uparrow+P_\downarrow,P_d\right)$ and the spin
expectation value, in units of $\hbar$, $\mathbf{S}=\left(S_x,S_y,S_z\right)$,
whose components are given by 
\begin{equation}
S_x=\frac{P^{\uparrow}_{\downarrow}+P^{\downarrow}_{\uparrow}}{2};\
S_y=i\frac{P^{\uparrow}_{\downarrow}-P^{\downarrow}_{\uparrow}}{2};\ 
S_z=\frac{P_{\uparrow}-P_{\downarrow}}{2}.
\end{equation}
Since we consider spin-degenerate dot levels, $E_\uparrow=E_\downarrow$, 
we can drop the first term on the r.h.s. of Eq.~(\ref{master}).

We are concerned with adiabatic pumping, where no transport voltage is applied
across the dot. The leads are therefore described by the same Fermi function
$f(\omega)$.  
As a consequence, the instantaneous current through the dot vanishes, 
and we need to consider the first adiabatic correction in order to
obtain the pumping current through the dot.
We  perform an adiabatic expansion along the lines of
Ref. \onlinecite{splett06}.  
We start by performing a Taylor expansion around the time $t$ of
$\mathbf{P}(t')$ appearing inside the integral on the r.h.s. of the
generalized master equation
\begin{eqnarray}
\frac{d}{dt}P^{\chi_1}_{\chi_2}\left(t\right) & = & \sum_{\chi_1',\chi_2'}
\int_{-\infty}^{t}dt'W^{\chi_1\chi_1'}_{\chi_2\chi_2'}\left(t,t'\right)\left[
P^{\chi_1'}_{\chi_2'}\left(t\right)\right.\nonumber\\
&& \left. + \left(t'-t\right)
\frac{d}{dt}P^{\chi_1'}_{\chi_2'}\left(t\right)\right]\ .
\end{eqnarray}
This expansion is justified by the fact that the response time of the system
is much smaller than the timescale of the parameter variation in time.
We then expand the kernel itself as
\begin{equation}
\label{wexp}
W^{\chi_1\chi_1'}_{\chi_2\chi_2'}\left(t,t'\right) \rightarrow 
\left(W^{\chi_1\chi_1'}_{\chi_2\chi_2'}\right)^{(i)}_t\left(t-t'\right)
+\left(W^{\chi_1\chi_1'}_{\chi_2\chi_2'}\right)^{(a)}_t\left(t-t'\right). 
\end{equation}
The superscript $i (a)$ denotes the instantaneous contribution (its adiabatic
correction). The instantaneous contribution corresponds to freezing all 
parameters to their values at time $t$, i.e. $X(\tau)\rightarrow X(t)$. The
adiabatic correction is obtained by linearizing the time dependence of the
parameters, i.e. $X(\tau)\rightarrow X(t)+(\tau-t)d/d\tau X(\tau)|_{\tau=t}$,
and retaining only first-order terms in the time derivatives.  
Finally, we perform the adiabatic expansion of the elements of the reduced
density matrix  
\begin{equation}
\label{pexp}
P^{\chi_1}_{\chi_2}\left(t\right)  \rightarrow 
\left(P^{\chi_1}_{\chi_2}\right)^{(i)}_t+
\left(P^{\chi_1}_{\chi_2}\right)^{(a)}_t\ .
\end{equation}
The subscript $t$ in Eqs. (\ref{wexp}) and (\ref{pexp})  denotes the time with
respect to which the adiabatic expansion is performed. This time $t$ enters
the respective quantities parametrically; both the instantaneous and the
adiabatic correction to the kernel are functions of the
time difference $(t-t')$. At this stage, it is convenient to introduce the
zero-frequency Laplace transform of the kernel as 
$
\left(W^{\chi_1\chi_1'}_{\chi_2\chi_2'}\right)^{(i/a)}_t=
\int_{-\infty}^tdt'\left(W^{\chi_1\chi_1'}_{\chi_2\chi_2'}\right)^{(i/a)}_t
(t-t') 
$. In order to evaluate the kernel of the master equation we perform, on top of
the adiabatic expansion, a perturbation expansion in the tunnel coupling
$\Gamma$. In the following we 
take into account processes in first order in the tunnel coupling. This
approach is valid in the weak-coupling limit, i.e.
$k_\mathrm{B}T\gg\Gamma$. At the same time the condition for adiabaticity,
$\Gamma\gg\Omega$, needs to be fulfilled. 
The instantaneous occupation probabilities and their adiabatic corrections
obey the equations  
\begin{eqnarray}
&&0=\sum_{\chi_1',\chi_2'}\left(W^{\chi_1\chi_1'}_{\chi_2\chi_2'}
\right)^{(i,1)}_t\left(P^{\chi_1'}_{\chi_2'}\right)^{(i,0)}_t\\
&&\frac{d}{dt}\left(P^{\chi_1}_{\chi_2}\right)^{(i,0)}_t
=\sum_{\chi_1',\chi_2'}\left(W^{\chi_1\chi_1'}_{\chi_2\chi_2'}
\right)^{(i,1)}_t\left(P^{\chi_1'}_{\chi_2'}\right)^{(a,-1)}_t .
\end{eqnarray}
The number in the superscripts designates the order in the perturbation
expansion in the tunnel coupling. The fact that we find elements of the
reduced density matrix in minus first order in the tunnel coupling is
consistent with our perturbative scheme, as those terms are proportional to
$\Omega/\Gamma$, which is small in the adiabatic limit. The
evaluation of the matrix elements of the kernel is done using a real-time
diagrammatic technique, which was developed in Ref. \onlinecite{koenig96},
extended to systems containing ferromagnetic leads in
Ref. \onlinecite{braun04}, and extended to adiabatic pumping in
Ref. \onlinecite{splett06}. The adiabatic correction to the matrix elements of
the kernel does not appear in lowest order in the tunnel coupling, as
considered in this paper. The equations for the instantaneous probabilities
and for the adiabatic correction to the probabilities can be summarized as  
\begin{widetext} 
\begin{eqnarray}
&&\hbar\frac{d}{dt}\left(\begin{array}{c} P_0\\ P_1\\ P_d\end{array}\right)
 = \label{eq_master_prob}\\
&& \Gamma\left(
\begin{array}{ccc}
-2 f(\epsilon) & 1-f(\epsilon) & 0\\
2 f(\epsilon) & -\left[1-f(\epsilon)+f(\epsilon+U)\right]&
2\left[1-f(\epsilon+U)\right]\\
0 & f(\epsilon+U) & -2\left[1-f(\epsilon+U)\right]
\end{array}
\right)
\left(\begin{array}{c} P_0\\ P_1\\ P_d\end{array}\right)
+
\left(
\begin{array}{c}
1-f(\epsilon)\\
-\left[1-f(\epsilon)-f(\epsilon+U)\right]\\
-f(\epsilon+U)
\end{array}
\right)\sum_\alpha 2\Gamma_\alpha
\mathbf{S}\cdot\mathbf{p}_\alpha\nonumber\ .
\end{eqnarray}
Similarly, the equations for the expectation value of the spin read
\begin{eqnarray}
\hbar \frac{d}{dt}
\mathbf{S}
&  = &
\left[
f(\epsilon)P_0
-\frac{1}{2}\left[
             1-f(\epsilon)-f(\epsilon+U)
            \right]P_1
-\left[1-f(\epsilon+U)\right]P_d
\right]
\sum_\alpha \Gamma_\alpha \mathbf{p}_\alpha
\nonumber\\
 & - &\Gamma\left[1-f(\epsilon)+f(\epsilon+U)\right]
\mathbf{S}
+\mathbf{S}\times\sum_\alpha\mathbf{B}_\alpha\label{eq_master_spin}\ ,
\end{eqnarray}
\end{widetext}
where we introduced the notation 
$\mathbf{p}_\alpha =  p_\alpha\mathbf{\hat{n}}_\alpha$.
The interaction-induced exchange field or effective $B$-field appearing in
Eq.~(\ref{eq_master_spin}) is given by the principal-value integral
\begin{equation}\label{eq_bfield}
\mathbf{B}_\alpha=\Gamma_\alpha \mathbf{p}_\alpha\int_P\frac{d\omega}{\pi}
\left(\frac{1-f(\omega)}{\omega-\epsilon}+
\frac{f(\omega)}{\omega-\epsilon-U}\right)\ .
\end{equation}
The instantaneous elements of the reduced density matrix are obtained by
setting the left hand side of Eqs. (\ref{eq_master_prob}) and
(\ref{eq_master_spin}) to zero and by assigning superscripts $(i,0)$ to the
vectors $\mathbf{P}$ and $\mathbf{S}$ on the right hand side of the
equations. The first adiabatic corrections to the
elements of the reduced density matrix are obtained by assigning to the
vectors $\mathbf{P}$ and $\mathbf{S}$ the superscripts $(i,0)$ on the left
hand side of the equations and the superscripts $(a,-1)$ on the right hand
side.  

\section{Results}

Starting from the master equation we compute both the instantaneous matrix
elements of the reduced density matrix and their first adiabatic correction.
These are needed as an input for calculating the spin and charge current.
\subsection{Dot occupation and spin}\label{sec_occup}

The fact that no transport voltage is applied to the system has important
consequences on the occupation probabilities and the expectation value of
the spin. 
To lowest order in the tunnel-coupling strength $\Gamma$, the 
instantaneous occupation probabilities are given by their equilibrium values, 
i.e. by the Boltzmann factors of the respective states 
\begin{eqnarray}
P_\chi^{(i,0)} & = &\frac{e^{-\beta E_\chi(t)}}{Z}\ ,
\end{eqnarray}
where $\beta=1/k_\mathrm{B}T$ is the inverse temperature, $E_\chi(t)$  the
energy of the dot state $\chi$, and $Z$ the partition function.
The spin expectation value vanishes,\cite{comment_equilibrium}
\begin{equation}
\mathbf{S}^{(i,0)}=0\ .
\end{equation}

When considering only first-order tunneling processes, the adiabatic
correction to the reduced density matrix is linear in $\Omega/\Gamma$. 
While the spin polarization of the leads has no influence on
the instantaneous probabilities, the situation is different for the adiabatic
correction, which reads    
\begin{equation}
\label{eq_prob}
\mathbf{P}^{(a,-1)}  =  -\frac{d\mathbf{P}^{(i,0)}}{dt}
\tau_{\mathrm{rel}}^Q(t)\frac{\Gamma(t)^2 }{\Gamma^2(t)-
\left(\sum_{\alpha}\mathbf{p}_\alpha\Gamma_\alpha(t)\right)^2}\ ,
\end{equation}
with the charge relaxation time given by
$\tau_{\mathrm{rel}}^Q(t)$ with 
$\left(\tau_{\mathrm{rel}}^Q(t)\right)^{-1} = \Gamma(t)
\left[1+f(\epsilon(t))-f(\epsilon(t)+U)\right]$ (the derivation of the
expression for $\tau_{\mathrm{rel}}^Q$ is given in Appendix \ref{appA}.)   
For vanishing polarization in the two leads this result coincides 
with that obtained for an N-dot-N system\cite{splett06,comment_factor2}  
\begin{equation}
\mathbf{P}^{(a,-1)}=-\frac{d\mathbf{P}^{(i,0)}}{dt}\frac{1}{\Gamma(t)}
\frac{1}{1+f(\epsilon(t))-f(\epsilon(t)+U)}\ .
\end{equation}
We find non-vanishing contributions to the off-diagonal terms of the reduced
density matrix, which vanish for zero polarization in the leads.
The spin expectation value reads
\begin{eqnarray}\label{eq_spin}
\mathbf{S}^{(a,-1)} =  \frac{1}{2}
\frac{\partial\langle n\rangle^{(i,0)}(t)}{\partial t}\tau_{\mathrm{rel}}^S(t)
\frac{\Gamma(t)\sum_{\alpha}\mathbf{p}_\alpha\Gamma_\alpha(t)}
{\Gamma^2(t)-\left(\sum_{\alpha}\mathbf{p}_\alpha\Gamma_\alpha(t)\right)^2}\ ,
\end{eqnarray}
where the spin relaxation time $\tau_{\mathrm{rel}}^S(t)$ is given by 
$\left(\tau_{\mathrm{rel}}^S(t)\right)^{-1}= \Gamma(t) \left[1-f(\epsilon(t))+
f(\epsilon(t)+U)\right]$ (the derivation of the expression for 
$\tau_{\mathrm{rel}}^S$ is given in Appendix \ref{appA}). 
Notice that the first adiabatic correction is the leading contribution to the
expectation value of the dot spin. Furthermore, a time-dependent dot spin can
be accumulated only by varying in time the occupation of the
dot.
The adiabatic correction to the spin component is parallel to the 
exchange field, introduced in Eq.~(\ref{eq_bfield}). Therefore no
precession of the spin around this field takes place. This is
different from the case of a time-independent but biased spin 
valve.\cite{braun04}

Finally, we remark that the limit where both leads are fully polarized
along the same magnetization axis, $\phi=0$ and $p_\mathrm{L}= p_\mathrm{R}=1$
is ill defined; in fact, in this case the life time of a minority spin in the
dot diverges and consequently in order for the adiabatic expansion to hold the
pumping frequency needs to be zero.

\subsection{Pumping Current}
 
The results for the dot occupation probabilities and the expectation value of
the spin on the dot serve  to calculate the pumping current.  
Using a similar approach as for the generalized master equation we write
the current into the left lead as
\begin{equation}
I_\mathrm{L}(t)=-e\int_{-\infty}^{t}dt'\sum_{\chi_1,\chi_2,\chi_1',\chi_2'}
\left(W_{\chi_1\chi_1'}^{\chi_2\chi_2'}\right)^\mathrm{L}(t,t')
P_{\chi_1'}^{\chi_2'}(t')\ ,
\end{equation}
where $\left(W_{\chi_1\chi_1'}^{\chi_2\chi_2'}\right)^\mathrm{L}(t,t')
=\sum_qq\left(W_{\chi_1\chi_1'}^{\chi_2\chi_2'}\right)^{\mathrm{L}q}(t,t')$,
and $\left(W_{\chi_1\chi_1'}^{\chi_2\chi_2'}\right)^{\mathrm{L}q}(t,t')$ is the
sum of all processes, describing transitions where the difference of the
number of electrons entering and leaving the left lead is equal to the
integer number $q$.

We compute the first order adiabatic correction to the current including only
first-order tunneling processes. We find 
\begin{eqnarray}
I_\mathrm{L}^{(a,0)}(t) & = & -e\sum_{\chi_1,\chi_2,\chi_1',\chi_2'}
\left(W_{\chi_1\chi_1'}^{\chi_2\chi_2'}\right)^{\mathrm{L}(i,1)}_t
\left(P_{\chi_1'}^{\chi_2'}\right)^{(a,-1)}_t\ .\nonumber
\end{eqnarray}
In the following we suppress the superscript $(a,0)$ for the current, since
the instantaneous current is always zero and $I_\mathrm{L}^{(a,0)}(t)$ is
therefore the dominant contribution. The current is of zeroth order in the
tunnel coupling and proportional to the pumping frequency $\Omega$.
To this order in the tunnel-coupling strengths, the pumped current is
non-vanishing only if the dot level position is one of the pumping parameters,
since $\left(P_{\chi_1'}^{\chi_2'}\right)^{(a,-1)}$ is proportional to the
time derivative of the dot level position. 
We find for the pumping current
\begin{eqnarray}
I_\mathrm{L} & = & 
-e\frac{\partial\langle n\rangle^{(i,0)}}{\partial t}
\frac{\Gamma_\mathrm{L}(t)}{\Gamma(t)} \nonumber\\
& - & e\frac{\partial\langle
n\rangle^{(i,0)}}{\partial
t}\frac{\Gamma_\mathrm{L}(t)\Gamma_\mathrm{R}(t)}{\Gamma(t)^2}
\frac{\left(\mathbf{p}_\mathrm{R}-\mathbf{p}_\mathrm{L}\right)
\boldsymbol{\pi}(t)}{1-\boldsymbol{\pi}(t)^2}
 \ ,\label{eq_charge_current_f}
\end{eqnarray}
where we have defined  the quantity
$\boldsymbol{\pi}(t)  =  \sum_\alpha\Gamma_\alpha(t)
\mathbf{p}_\alpha/\Gamma(t)=\sum_\alpha\boldsymbol{\pi}_\alpha(t)$, 
which depends on time via $\Gamma_\alpha(t)$.
The pumped current consists of two terms of different origin: the
first one is independent of the lead polarizations, and can be interpreted as
arising from a peristaltic mechanism;\cite{splett06} the second term depends
on the  polarizations and can be seen as arising from the relaxation of the
accumulated spin on the dot.  The latter contribution due to spin relaxation can be either positive or
negative depending on the polarization strengths, polarization directions and
tunnel coupling to the different leads. This means that the time-resolved
current can be enhanced with respect to the non-magnetic case, in contrast
with the spin-valve effect in a time-independent system, which always leads to
a current \textit{suppression}. 
Similarly, we will see later that also the charge pumped
through the dot per period is always suppressed due to the polarization of the
leads. Therefore, this inverse spin-valve effect is observable only in the
time-resolved current response. The effect could be investigated
experimentally  by means of time-resolved measurements or by rectifying the
current response. Reduction or enhancement of the current can be achieved by
tuning the tunnel coupling or the lead magnetizations.

\subsection{N-dot-F: Spin Pumping}

We now turn our attention to spin pumping in a setup where only one of the
leads, the right one for the sake of definiteness, is ferromagnetic. 
We calculate the spin pumped in the unpolarized left lead. 

The instantaneous contribution to 
the reduced density matrix is independent of the polarization and therefore
remains unchanged. The first adiabatic correction for the occupation
probabilities and the dot spin are given by Eq.~(\ref{eq_prob}) and
Eq.~(\ref{eq_spin}), respectively,  with $p_\mathrm{L}=0$. 

The charge current through such a N-dot-F system can be
obtained directly from Eq. (\ref{eq_charge_current_f}) by setting
$p_\mathrm{L}=0$ and it reads 
\begin{equation}\label{eq_charge_current}
I_\mathrm{L} = -e\frac{\partial\langle n\rangle^{(i,0)}}{\partial t}
\frac{\Gamma(t)\Gamma_{\mathrm{L}}(t)}{\Gamma(t)^2-p_\mathrm{R}^2
\Gamma_\mathrm{R}(t)^2}\ .
\end{equation}

For calculating the spin current, we chose a global spin-quantization axis
parallel to the magnetization of the right lead. In this basis the reduced
density matrix does not have any off-diagonal terms. We find for the spin
current 
\begin{equation}\label{eq_spin_current}
I_\mathrm{L}^{S} = \frac{\hbar}{2}\frac{\partial\langle n\rangle^{(i,0)}}
{\partial t}
\frac{\Gamma_\mathrm{R}(t)p_\mathrm{R}\Gamma_{\mathrm{L}}(t)}{\Gamma(t)^2-
p_\mathrm{R}^2\Gamma_\mathrm{R}(t)^2}\ .
\end{equation}
The ratio of the time-resolved spin and charge currents,
Eq.~(\ref{eq_charge_current}) and Eq.~(\ref{eq_spin_current}), reads  
\begin{eqnarray}
\label{sqfact}
\frac{I_\mathrm{L}^{S}}{I_\mathrm{L}}/\left[\frac{\hbar/2}{-e}\right]
& = &
\frac{p_\mathrm{R}\Gamma_\mathrm{R}(t)}{\Gamma(t)}.
\end{eqnarray}
This ratio is, in general, time dependent. The time-resolved spin current is
smaller than the time-resolved particle current at any time $t$. The 
ratio of these two currents is always positive, implying that spin and charge
flow in the same direction, as expected. 
The situation is different, for the spin (in units of $\hbar/2$) and
the charge (in units of $-e$) \textit{pumped per period}. 
We calculate the pumped charge and spin for the following two choices of
pumping parameters, $\left\{\Gamma_\mathrm{L},\epsilon\right\}$ or
$\left\{\Gamma_\mathrm{R},\epsilon\right\}$, in bilinear response, i.e. we
calculate the pumped charge and spin per infinitesimal area in parameter
space.  
\begin{figure}
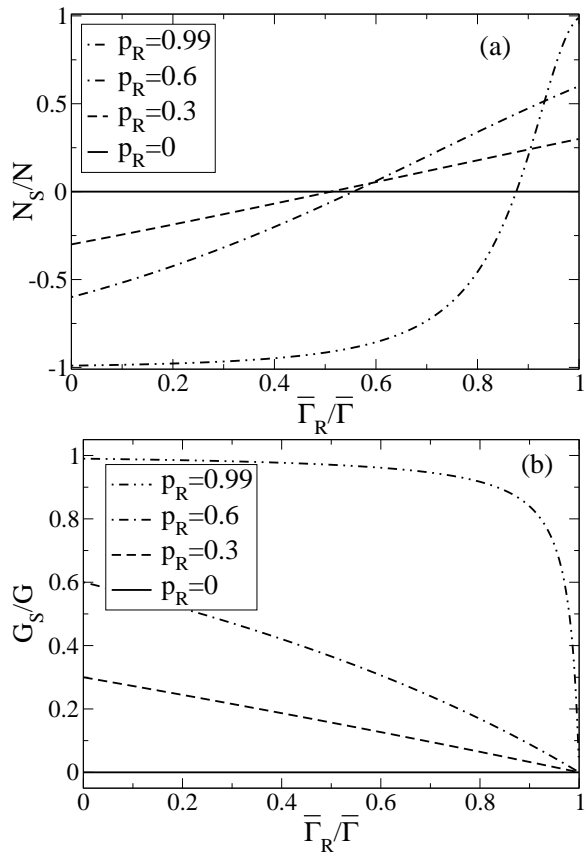

\begin{center}
\includegraphics[width=3.in]{fig2a}
\includegraphics[width=3.in]{fig2b}
\caption{
(a) Ratio of the pumped spin per period (in units of $\hbar/2$) to the
   pumped charge per period (in units of $-e$) as a function of the relative
   tunnel-coupling strength $\bar{\Gamma}_\mathrm{R}/\bar{\Gamma}$, for
   different polarizations of the right lead. 
(b)  Ratio between the linear dc spin conductance and the linear dc
   conductance as a function of the relative tunnel-coupling strength
   $\bar{\Gamma}_\mathrm{R}/\bar{\Gamma}$, for different polarizations of the
   right lead.}  
\label{fig_ratio_p}
\end{center}
\end{figure}
The result for the ratio of the
pumped spin (in units of $\hbar/2$) per period, $N_S$, and the 
pumped charge  (in units of $-e$) per period, $N$, reads
\begin{eqnarray}
\frac{N_S}{N}
\label{eq_charge_ratio}
 & = &
-p_\mathrm{R}\frac{
1+p_\mathrm{R}^2\left( \bar{\Gamma}_\mathrm{R} / \bar{\Gamma}\right)^2
-2\left( \bar{\Gamma}_\mathrm{R} / \bar{\Gamma}\right)
}{
1+p_\mathrm{R}^2\left( \bar{\Gamma}_\mathrm{R} / \bar{\Gamma}\right)^2-
2\left( \bar{\Gamma}_\mathrm{R} / \bar{\Gamma} \right)p_\mathrm{R}^2
}\ .
\end{eqnarray}
It turns out that the efficiency of the spin pump does not depend on which
pair of pumping parameters one chooses. In Fig. \ref{fig_ratio_p}(a), we plot
the ratio of pumped spin to pumped charge as a function of the relative
tunnel-coupling strength ${\bar{\Gamma}_\mathrm{R}}/{\bar{\Gamma}}$, where the
bar indicates time-averaged quantities, for different values of the
polarization of the right lead.  
The absolute value of the ratio is maximally equal to one in the case of
full polarization of the right lead. 
For $p_{\mathrm{R}}<1$, this ratio goes from $-p_{\mathrm{R}}$ for vanishing
$\bar{\Gamma}_{\mathrm{R}}$ to $p_{\mathrm{R}}$ for vanishing
$\bar{\Gamma}_{\mathrm{L}}$ 
changing its sign for 
\begin{equation}
\frac{\Gamma_\mathrm{R}}{\Gamma}=\frac{1}{p_\mathrm{R}^2}
\left(1-\sqrt{1-p_\mathrm{R}^2}\right)\ .
\end{equation}
This is a very intriguing result, which implies that the respective direction
in which spin and charge are pumped depends on the coupling to left and right
lead. 

\begin{figure}[t]
\begin{center}
\includegraphics[width=3.in]{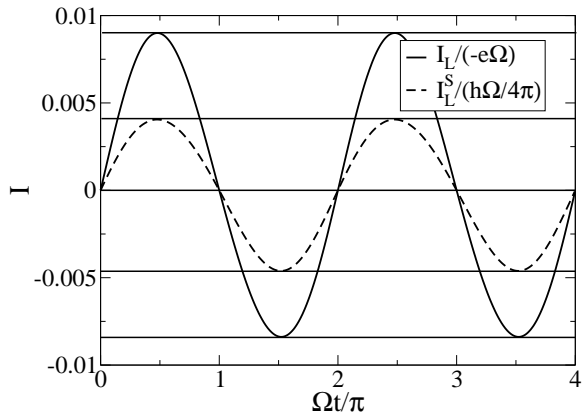}
\end{center}
\caption{Time-resolved spin and charge currents as a function of time. The
  value of the parameters used for this plot are:
  $p_\mathrm{R}=0.99$,$\bar{\Gamma}_\mathrm{L}=\bar{\Gamma}_\mathrm{R}$, 
$\bar{\epsilon}=-\bar{\Gamma}$, $U=10\bar{\Gamma}$,
$|\delta\Gamma_\mathrm{L}|/\bar{\Gamma}=|\delta\epsilon|/\bar{\Gamma}=0.1$ and
$k_{\mathrm{B}}T=\bar{\Gamma}$.}
\label{fig_timedep}
\end{figure}

The average pumped charge and the average pumped spin can
have opposite signs, while the time-resolved spin and charge currents flow in
the same direction at any instant of time, due to the fact that the ratio of
the time-resolved currents, Eq.~(\ref{sqfact}), is itself time dependent. 
To elucidate this, in Fig. \ref{fig_timedep}, we plot the time-resolved spin
and charge currents as a function of time for a configuration, where the
pumped spin and charge per period have  different signs. Note that the charge
current has a positive average and the spin current has a negative average,
while both currents flow in the same direction at any time.

We now contrast the results for the pumped spin and charge with the dc
transport properties of the N-dot-F system. We find for the spin and charge
currents 
\begin{widetext}
\begin{equation}
\frac{I^{S}}{I}/\left[\frac{\hbar/2}{-e}\right]= 
\frac{[1-f_\mathrm{L}(\epsilon)+f_\mathrm{L}(\epsilon+U)]
\Gamma_\mathrm{L}p_\mathrm{R}}
{[1-f_\mathrm{L}(\epsilon)+f_\mathrm{L}(\epsilon+U)]\Gamma_\mathrm{L}+
[1-f_\mathrm{R}(\epsilon)+f_\mathrm{R}(\epsilon+U)]
\Gamma_\mathrm{R}(1-p_\mathrm{R}^2)}\ , 
\end{equation}
\end{widetext}
which, in the linear response regime, yields for the ratio of the spin to the 
charge conductance 
\begin{equation}
\frac{G^{S}}{G}/\left[\frac{\hbar/2}{-e}\right]=
\frac{\Gamma_\mathrm{L}p_\mathrm{R}}{\Gamma_\mathrm{L}+\Gamma_\mathrm{R}
(1-p_\mathrm{R}^2)}\ .
\end{equation}
The linear conductance ratio is shown in Fig.~\ref{fig_ratio_p}(b). 
Its behavior is completely different from that obtained by pumping.
First, the spin polarization decreases as a function of 
$\bar \Gamma_\mathrm{R} / \bar \Gamma$ and, second, it stays always positive.

Finally, we consider the spin accumulated on the dot in one
pumping period. We find two
different results depending on whether
$\left\{\Gamma_\mathrm{L},\epsilon\right\}$ or
$\left\{\Gamma_\mathrm{R},\epsilon\right\}$ are the pumping 
parameters: 
\begin{eqnarray}
&&\langle
\mathbf{S}^{(a,-1)}\rangle_T\left(\left\{\Gamma_\mathrm{L},\epsilon\right\}
\right)
 =  -\frac{\eta}{\bar{\Gamma}}\frac{\partial\langle
\bar{n}\rangle}{\partial\bar{\epsilon}}\bar{\tau}^S_{\mathrm{rel}}
\frac{\bar{\boldsymbol{\pi}}_\mathrm{R}}{(1-\bar{
\boldsymbol{\pi}}_\mathrm{R}^2)^2}\\ 
&&\langle
\mathbf{S}^{(a,-1)}\rangle_T
\left(\left\{\Gamma_\mathrm{R},\epsilon\right\}\right)
 =  \frac{\eta}{2\bar{\Gamma}}\frac{\partial\langle
\bar{n}\rangle}{\partial\bar{\epsilon}}\bar{\tau}^S_{\mathrm{rel}}
\frac{(\frac{\bar{\Gamma}_\mathrm{L}}{\bar{\Gamma}}-
\frac{\bar{\Gamma}_\mathrm{R}}{\bar{\Gamma}})
+\bar{\boldsymbol{\pi}}_\mathrm{R}^2} 
{(1-\bar{\boldsymbol{\pi}}_\mathrm{R}^2)^2}
\mathbf{p}_\mathrm{R}\nonumber\ ,\\
\end{eqnarray}
where the area of the cycle in parameter space, $\eta$, is defined
as $\eta=\int_0^Tdt\frac{\partial\epsilon}{\partial t}\delta\Gamma_\mathrm{L}$
and 
$\eta=\int_0^Tdt\frac{\partial\epsilon}{\partial t}\delta\Gamma_\mathrm{R}$
for the first and second equation respectively.  
In the case of pumping with $\left\{\Gamma_\mathrm{R},\epsilon\right\}$,
i.e. when the coupling to the ferromagnetic lead is time dependent, the
average spin changes sign at the same values of
$\bar{\Gamma}_\mathrm{R}/\bar{\Gamma}$ at which the the ratio
$I_{\mathrm{L}}^{\mathrm{S}}/I^{}_{\mathrm{L}}$ changes its sign. 
On the contrary,
when pumping with $\Gamma_{\mathrm{L}}$ and $\epsilon$, the average spin
polarization of the dot does not change sign as a function of 
$\bar{\Gamma}_\mathrm{R}/\bar{\Gamma}$, while the ratio
$I_{\mathrm{L}}^{\mathrm{S}}/I^{}_{\mathrm{L}}$ still does.

\subsection{F-dot-F: Spin-Valve Effect}
We now consider the spin-valve setup with both leads having arbitrary spin
polarizations. We compute the number of pumped charges per period,  
$ N = -\frac{1}{e}\int_0^TdtI_\mathrm{L}(t)$, in bilinear response in the
pumping parameters. For the pumping cycle defined by
$\epsilon(t)=\bar{\epsilon}+\delta\epsilon(t)$ and
$\Gamma_{\mathrm{L}}(t)=\bar{\Gamma}_{\mathrm{L}}
+\delta\Gamma_{\mathrm{L}}(t)$,
the number of pumped charges per period reads
\begin{eqnarray}
 N & = & \label{eq_chargenumber}
\eta\frac{\partial\langle
  \bar{n}\rangle^{(i,0)}}{\partial\bar{\epsilon}}\\
 && \frac{\partial}{\partial
  \bar{\Gamma}_\mathrm{L}}\left(
\frac{\bar{\Gamma}_\mathrm{L}\sum_{\alpha}\bar{\Gamma}_\alpha
-\bar{\Gamma}_\mathrm{L}\mathbf{p}_\mathrm{L}\sum_\alpha\bar{\Gamma}_\alpha
\mathbf{p}_\alpha }{
\bar{\Gamma}^2-\left(\sum_\alpha\bar{\Gamma}_\alpha
\mathbf{p}_\alpha\right)^2}
\right)\nonumber\ ,
\end{eqnarray}
with $\eta=\int_0^Tdt\frac{\partial\epsilon}{\partial
  t}\delta\Gamma_\mathrm{L}$ being the area of the cycle in parameter space.
Notice that the charge number in Eq. (\ref{eq_chargenumber}) is a product of
  two terms, where one contains effects of interactions and another one
  effects of the leads' magnetization.

\begin{figure}
\begin{center}
\includegraphics[width=3.in]{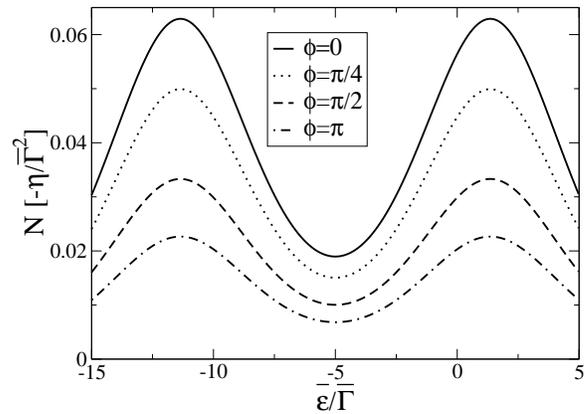}
\caption{Pumped charge as a function of the average level
position, $\bar{\epsilon}$, for different values of the angle between the
magnetizations. The polarizations in the leads are
$p_\mathrm{L}=p_\mathrm{R}=0.8$.} 
\label{fig_charge_eps}
\end{center}
\end{figure}

\begin{figure}
\begin{center}
\includegraphics[width=3.in]{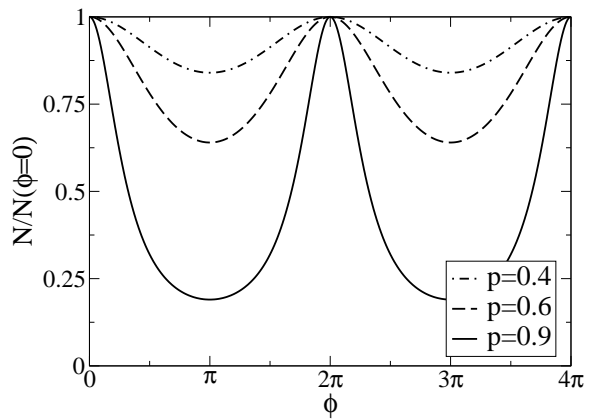}
\caption{
Pumped charge as a function of the angle between the
magnetizations of the leads, $\phi$, for different polarization strengths
$p_\mathrm{L}=p_\mathrm{R}=p$. This result does not depend on the level
position and the interaction strength.
  }
\label{fig_charge_phi}
\end{center}
\end{figure}

In the following we show results for the case that both leads have the same
spin polarization strength. This corresponds to the experimentally relevant
situation that both leads are realized with the same ferromagnetic material. In
Fig.~\ref{fig_charge_eps}, we show  the pumped charge as a function of the
level position for different values of the angle between the directions of the
magnetizations of the two leads. The pumped charge shows  a peak  when the
energy $\bar{\epsilon}$ or  $\bar{\epsilon}+U$ are close to the Fermi energy,
similarly to pumping through a quantum dot contacted to two non-magnetic leads
(N-dot-N).\cite{splett06} As far as the dependence on the angle between the
magnetization of the two leads $\phi$ is concerned, for $\phi\in[0,\pi]$, the
charge is monotonically suppressed for increasing $\phi$ until a minimum is
reached for $\phi=\pi$ as in the usual dc spin-valve effect. The full
$\phi$-dependence of the pumped charge is shown in Fig. \ref{fig_charge_phi},
where we plot $N(\phi)/N(\phi=0)$.   
This result does not depend on the  value of the level position and of the
interaction strength, since the dependence on $\bar{\epsilon}$ and $U$ cancels
out when we divide by $N(\phi=0)$. We notice that the suppression of charge
pumping is stronger for higher lead polarizations. 
Furthermore, the more the lead polarization is increasing the stronger the
behavior of the pumped charge as a function of the angle deviates from a
cosine law. 

\begin{figure}
\begin{center}
\includegraphics[width=3.in]{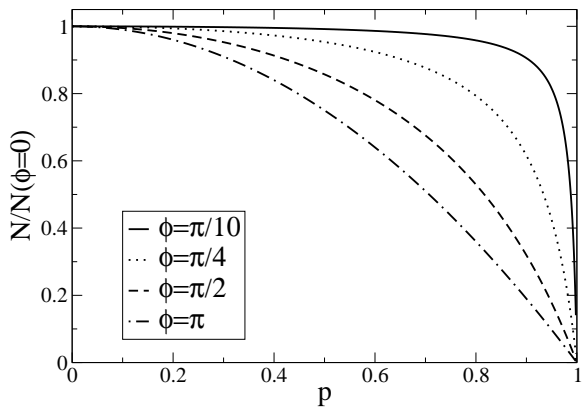}
\caption{Pumped charge as a function of the polarization strength
  $p=p_\mathrm{L}=p_\mathrm{R}$, for different values of the angle between the
  magnetizations. This result does not depend on the level position and the
  interaction strength. 
  }
\label{fig_charge_p}
\end{center}
\end{figure}
In Fig. \ref{fig_charge_p}, we show the pumped charge as a function
of the lead polarization strengths for different values of the angle
between the magnetization directions. This plot confirms that the pumped charge
decreases for increasing spin polarization of the leads. 
The charge suppression is strongest when $\phi$ is near to $\pi$. 
Independently of the angle between the polarization axis of
 left and right lead, the pumped charge goes to zero for fully polarized
leads. 
It is important to point out that this last property depends on 
the order in which limits are taken, since  the two limits $\phi\rightarrow 0$
and $p_\mathrm{L}=p_\mathrm{R}\rightarrow 1$ do not commute, as was already
mentioned in section \ref{sec_occup}. In fact, 
comparing Fig. \ref{fig_charge_phi} and Fig. \ref{fig_charge_p} we
notice that in the first case the charge is maximal for $\phi=0$
even for the polarization increasing towards one, while in the
second case for $p_\mathrm{L}=p_\mathrm{R}=1$,  the charge is
maximally suppressed even for $\phi$ going to zero.

The spin, which is accumulated on the dot 
during one pumping cycle is given by 
\begin{eqnarray}\label{eq_average_spin}
\langle \mathbf{S}\rangle^{(a,-1)}_{T} & = &
\frac{\eta}{2}\frac{\partial\langle
  \bar{n}\rangle^{(i,0)}}{\partial\epsilon}\bar{\tau}_\mathrm{rel}^S
\left[\frac{\bar{\Gamma}\mathbf{p}_\mathrm{L}}{\bar{\Gamma}^2-
\left(\sum_\alpha \mathbf{p}_\alpha \bar{\Gamma}_\alpha\right)^2}\right.\\
&& \left.-2\frac{ \bar{\Gamma}^2-\bar{\Gamma}\mathbf{p}_\mathrm{L}\sum_\alpha
    \mathbf{p}_\alpha \bar{\Gamma}_\alpha}{\left(\bar{\Gamma}^2-
\left(\sum_\alpha  \mathbf{p}_\alpha \bar{\Gamma}_\alpha\right)^2\right)^2}
\cdot \sum_\alpha
\mathbf{p}_\alpha \bar{\Gamma}_\alpha\right]\nonumber, 
\end{eqnarray}
where the pumping parameters are chosen to be $\Gamma_\mathrm{L}$ and
$\epsilon$. The result for pumping with $\Gamma_\mathrm{R}$ and
$\epsilon$ is easily obtained by swapping the indices L and R. 
Depending on the spin polarization of the leads and on the values of
tunnel-coupling strengths, the average spin on the dot can point along any
direction in the plane containing the magnetizations of the leads.  

\section{Conclusions}

We have investigated adiabatic pumping through a single-level quantum dot with
ferromagnetic leads in the regime of weak tunnel coupling between dot and 
leads, by means of a real-time diagrammatic approach. In the case that only 
one lead is
ferromagnetic, we have computed the spin injected in the
non-magnetic lead by pumping. We have found that, depending on the relative
strength of the tunnel coupling to the leads, spin and charge can be pumped,
on average, in opposite directions.
For the case when both leads are polarized, we have found a suppression of 
the pumped charge by means of the spin-valve effect and determined the average
spin accumulated on the dot during one pumping cycle.

\acknowledgments

We would like to thank M. B\"uttiker for useful discussions. 
We acknowledge financial support from the EU via the STREP project SUBTLE, and
from the DFG via the SPP 1285 and the SFB 491.

\begin{appendix}

\section{Relaxation times}
\label{appA}

In this appendix, we calculate the spin and charge
relaxation times. In order to
calculate the spin relaxation time, we consider the case when the
charge on the dot is in equilibrium and the occupation
probabilities are therefore given by the Boltzmann factors. Then
Eq. (\ref{eq_master_spin}) simplifies to
\begin{equation}
\frac{d\mathbf{S}}{dt}=-\Gamma
\left[1-f(\epsilon)+f(\epsilon+U)\right]\mathbf{S}\ ,
\end{equation}
where we also made use of  the fact that the spin is always
parallel to the exchange field. The spin relaxation time is
therefore given by
\begin{equation}
\tau_\mathrm{rel}^{S}=\frac{1}{\Gamma}\frac{1}{1-f(\epsilon)+f(\epsilon+U)}\
.
\end{equation}
In order to calculate the charge relaxation time, we consider Eq.
(\ref{eq_master_prob}), where we take the spin in equilibrium,
such that $\mathbf{S}=0$. Then, we find for the dot occupation number
\begin{eqnarray}
\frac{d\langle
n\rangle}{dt} & = & \Gamma\left[2f(\epsilon)P_0-
(1-f(\epsilon)-f(\epsilon+U))P_1\right.\nonumber
\\ &  &- \left.2(1-f(\epsilon+U))P_d\right]\
.
\end{eqnarray}
Taking into account that the sum over the occupation probabilities
has to be equal to one at any instant in time, we find
\begin{equation}
\frac{d\langle
n\rangle}{dt} = -\Gamma\left[1+f(\epsilon)-f(\epsilon+U)\right]\left(\langle
n\rangle -\langle n\rangle_\mathrm{eq}\right)\ ,
\end{equation}
where $\langle n\rangle_\mathrm{eq}$ is the equilibrium occupation number of
the dot. The charge relaxation time is therefore given by
\begin{equation}
\tau_\mathrm{rel}^{Q}=\frac{1}{\Gamma}\frac{1}{1+f(\epsilon)-f(\epsilon+U)}\
.
\end{equation}
Both relaxation times depend strongly on the position of the dot level with
respect to the Fermi energy of the leads and on the strength of the Coulomb
interaction.

\end{appendix}


\begin{thebibliography}{24}





\bibitem{brouwer98} P. W. Brouwer, Phys. Rev. B {\bf 58}, R10135 (1998).

\bibitem{zhou99} F. Zhou, B. Spivak, and B. Altshuler, Phys. Rev. Lett.
        {\bf 82}, 608 (1999).

\bibitem{buttiker01} M. Moskalets and M. B\"uttiker, Phys. Rev. B {\bf 64},
        201305(R) (2001).

\bibitem{buttiker02} M. Moskalets and M. B\"uttiker, Phys. Rev. B {\bf 66},
        035306 (2002).

\bibitem{entin02} O. Entin-Wohlman, A. Aharony, and Y. Levinson,
        Phys. Rev. B {\bf 65}, 195411 (2002).




\bibitem{geerligs91} L. J. Geerligs, S. M. Verbrugh, P. Hadley, J. E. Mooij,
    H. Pothier, P. Lafarge, C. Urbina, D. Est\`eve, and M. H. Devoret,
    Z. Phys. B {\bf 85}, 349 (1991).

\bibitem{pothier92} H. Pothier, P. Lafarge, C. Urbina, D. Est\`eve, and
        M. H. Devoret, Europhys. Lett. {\bf 17}, 249 (1992).


\bibitem{switkes99} M. Switkes, C. M. Marcus, K. Campman, and A. C. Gossard,
        Science {\bf 283}, 1905 (1999).

\bibitem{fletcher03} N. E. Fletcher, J. Ebbecke, T. J. B. M. Janssen,
        F. J. Ahlers, M. Pepper, H. E. Beere, and D. A. Ritchie, Phys. Rev. B
        {\bf 68}, 245310 (2003); J. Ebbecke, N. E. Fletcher,
        T. J. B. M. Janssen, F. J. Ahlers, M. Pepper, H. E. Beere, and
        D. A. Ritchie, Appl. Phys. Lett. {\bf 84}, 4319 (2004).


\bibitem{watson03} S. K. Watson, R. M. Potok, C. M. Marcus, and V. Umansky,
                   Phys.   Rev. Lett {\bf 91}, 258301 (2003).


\bibitem{buttiker94} M. B\"uttiker, H. Thomas, and A. Pr\^etre, Z. Phys. B:
        Condens. Matter \textbf{94}, 133 (1994).


\bibitem{aleiner98} I. L. Aleiner and A. V. Andreev, Phys. Rev. Lett.
                      \textbf{81}, 1286 (1998).

\bibitem{citro03}  R. Citro, N. Andrei, and Q. Niu, Phys. Rev. B {\bf 68},
                     165312 (2003).

\bibitem{aono04}       T. Aono, Phys. Rev. Lett. \textbf{93}, 116601 (2004).


\bibitem{brouwer05} P. W. Brouwer, A. Lamacraft, and K. Flensberg,
               Phys. Rev. B  {\bf 72}, 075316 (2005).


\bibitem{cota05}  E. Cota, R. Aguado, and G. Platero, Phys. Rev. Lett. {\bf
                94}, 107202 (2005); E. Cota, R. Aguado, and G. Platero,
                Phys. Rev.  Lett. {\bf 94}, 229901(E) (2005).
\bibitem{splett05} J. Splettstoesser, M. Governale, J. K\"onig, and R. Fazio,
               Phys. Rev. Lett. \textbf{95}, 246803 (2005).

\bibitem{sela06} E. Sela and Y. Oreg, Phys. Rev. Lett. {\bf 96}, 166802 (2006).


\bibitem{splett06} J. Splettstoesser, M. Governale, J. K\"onig, and R. Fazio,
  Phys. Rev. B {\bf 74}, 085305 (2006).

\bibitem{fioretto07} D. Fioretto and A. Silva, arXiv:0707.3338 (2007).


\bibitem{braun04} 
J. K\"onig and J. Martinek, Phys. Rev. Lett. {\bf 90}, 166602 (2003);
M. Braun, J. K\"onig, and J. Martinek, Phys. Rev. B {\bf 70}, 195345 (2004).

\bibitem{QDSV}
J. Fransson, Europhys. Lett. {\bf 70}, 796 (2005); 
M. Braun, J. K\"onig, and J. Martinek, Europhys. Lett. {\bf 72}, 294
(2005);
I. Weymann and J. Barnas, Eur. Phys. J B {\bf 46}, 289 (2005);
S. Braig and P. W. Brouwer, Phys. Rev. B {\bf 71}, 195324 (2005);
W. Wetzels, G. E. W. Bauer, and M. Grifoni, Phys. Rev. B {\bf 72}, 020407(R)
(2005); 
J. N. Pedersen, J. Q. Thomassen, and K. Flensberg, Phys. Rev. B {\bf 72},
045341 (2005); 
M. Braun, J. K\"onig, and J. Martinek, Phys. Rev. B {\bf 74}, 075328 (2006);
I. Weymann and J. Barnas, Phys. Rev. B {\bf 75}, 155308 (2007);
D. Urban, M. Braun, and J. K\"onig, Phys. Rev. B {\bf 76}, 125306 (2007);
D. Matsubayashi and M. Eto, cond-mat/0607548;
R. P. Hornberger, S. Koller, G. Begemann, A. Donarini, and M. Grifoni,
arXiv:0712.0757. 

\bibitem{julliere75} M. Julli\`ere, Phys. Lett. {\bf 54A}, 225 (1975).

\bibitem{slonczewski} J. C. Slonczewski, Phys. Rev. B \textbf{39}, 6995 (1989).



\bibitem{mucciolo02}  E. R. Mucciolo, C. Chamon, and  C. M. Marcus,
  Phys. Rev. Lett.  {\bf 89}, 146802 (2002).

\bibitem{blaauboer03} M. Blaauboer, Phys. Rev. B {\bf 68}, 205316 (2003).

\bibitem{governale03}  M. Governale, F. Taddei, and R. Fazio, Phys. Rev. B {\bf
  68}, 155324 (2003).

\bibitem{zheng03} W. Zheng, J. Wu, B. Wang, J. Wang, Q. Sun, and H. Guo,
  Phys. Rev. B \textbf{68}, 113306 (2003). 

\bibitem{brataas02} A. Brataas, Y. Tserkovnyak, G. E. W. Bauer, and
  B. I. Halperin, Phys Rev. B \textbf{66},  060404(R), (2002). 

\bibitem{koenig96} J. K\"onig, H. Schoeller, and G. Sch\"on, Phys. Rev. Lett.
  {\bf76}, 1715 (1996);
  J. K\"onig, J. Schmid, H. Schoeller, and G. Sch\"on, Phys. Rev. B {\bf 54},
  16820 (1996);
  H. Schoeller, in \textit{Mesoscopic Electron Transport}, edited by L.L. Sohn,
  L.P. Kouwenhoven, and  G. Sch\"on (Kluwer, Dodrecht, 1997);
  J. K\"onig, \textit{Quantum Fluctuations in the Single-Electron Transistor}
  (Shaker, Aachen, 1999).



\bibitem{comment_equilibrium} This statement holds, also in the case of
  equally and fully polarized leads. The probability for a minority spin
  to enter the dot  goes to zero, but also its probability to leave the dot
  does, once the minority spin is on the dot.

\bibitem{comment_factor2}
Please note that the extra factor $1/2$ in the formula in 
Ref.~\onlinecite{splett06} is a misprint.
\end{thebibliography}
\end{document}